\documentclass[twocolumn,preprintnumbers,amssymb,amsmath,superscriptaddress,nofootinbib,groupedaddress]{revtex4}
\usepackage[dvips]{graphicx}
\usepackage{dcolumn} 
\usepackage{bm}
\usepackage{epsfig,amsmath,amssymb,verbatim,mathrsfs,array,layout,textcomp,amssymb,latexsym,slashed}
\usepackage{color}
\usepackage{hyperref}
\newcommand{\beq}{\begin{eqnarray}}
\newcommand{\eeq}{\end{eqnarray}}

\def\ltap{\ \raise.3ex\hbox{$<$\kern-.75em\lower1ex\hbox{$\sim$}}\ }
\def\gtap{\ \raise.3ex\hbox{$>$\kern-.75em\lower1ex\hbox{$\sim$}}\ }

\def\be{\begin{equation}}
\def\ee{\end{equation}}
\def\bea{\begin{eqnarray}}
\def\eea{\end{eqnarray}}

\pdfoutput=1

\def\ma5{{\tt MadAnalysis5\;}}

\def\ma5{{\tt MadAnalysis5\;}}

\begin{document} 


\title{
Boosting long lived particles searches at $\mu$TRISTAN}
\author{Daniele Barducci}
\email{daniele.barducci@pi.infn.it}
\affiliation{Universit\`a di Pisa and INFN Section of Pisa, Largo Bruno Pontecorvo 3, 56127, Pisa, Italia}
\vskip .05in
\begin{abstract}
\vskip .05in
We study the prospects of the proposed $\mu$TRISTAN experiment, running in the energy asymmetric $\mu^+ e^-$ mode, in probing long lived particles (LLPs) arising from the decay of the Standard Model Higgs boson. We focus on the proposed runs with $\{E_{\mu^+}, E_{e^-}\} = \{1\,{\rm TeV},\,30\,{\rm GeV}\}$ and $\{E_{\mu^+}, E_{e^-}\} = \{3\,{\rm TeV},\,50\,{\rm GeV}\}$ and we show that, owing to the boosted nature of the produced events, a far detector placed along the beam line can collect a large fraction of the LLP flux. This allows one to set bounds on the exotic Higgs branching ratio which, for specific $\phi$ decay modes, can surpass those expected at the end of the High Luminosity LHC in the regime of large LLPs proper decay lengths.
On the other hand, we find that the proposed strategy will not be able to further extend the limits that might be set by proposed LHC far detectors such as CODEX-b, ANUBIS and MATHUSLA.

\end{abstract}

\maketitle


\section{Introduction}\label{intro}

Light and weakly coupled dark sectors presenting new particles with masses in the ${\rm MeV}-{\rm GeV}$ range are commonly predicted in many New Physics (NP) theories that aim to explain the Standard Model (SM) shortcomings, such as the evidence for neutrino masses and Dark Matter. Given the absence of any manifestation of TeV scale NP at the LHC, these theories have nowadays become one of the main targets for present and future experimental efforts, see, {\emph{e.g.}},\,\cite{Antel:2023hkf} for a review.

Being light and weakly coupled, such new states typically have suppressed production rates and can exhibit macroscopic decay lengths, thus behaving as long lived particles (LLPs). At the LHC particles with small decay lengths of ${\cal O}(10^{-3}-10)\,$m can be searched for at the main purpose detectors ATLAS and CMS, by looking for displaced signatures. On the other hand, the existence of dark particles with longer lifetimes of ${\cal O}(10-100)\,$m can be directly tested by using detectors placed far away from the $pp$ interaction point (IP). This is for example the case of the currently operating FASER\,\cite{FASER:2018eoc,Feng:2017uoz} and SND@LHC\,\cite{SHiP:2020sos} and of proposed experiments as CODEX-b\,\cite{Gligorov:2017nwh,CODEX-b:2019jve,Dey:2019vyo}, ANUBIS\,\cite{Bauer:2019vqk,Brandt:2025fdj} and MATHUSLA\,\cite{Chou:2016lxi,Curtin:2018mvb,MATHUSLA:2018bqv,MATHUSLA:2025zyt}. Differently from searches performed at ATLAS and CMS, these far detectors usually work in an almost background free regime, owing to their large distance from the IP, but suffer from a small solid angle coverage,
as opposed to the almost $4\pi$ reach of ATLAS and CMS.
The search for LLPs is also one of the priorities when studying the potentiality of future colliders in testing light and weakly coupled dark sectors. Studies have tackled this problem for $e^+e^-$ colliders\,as FCC-ee, ILC and CEPC\,\cite{Antusch:2016vyf,Alipour-Fard:2018lsf,Wang:2019orr,Wang:2019xvx,Blondel:2022qqo,Ai:2025cpj}, hadron colliders as FCC-hh\,\cite{Bhattacherjee:2021rml,Boyarsky:2022epg,Bhattacherjee:2023plj,MammenAbraham:2024gun} and  $\mu^+\mu^-$colliders\,\cite{Alipour-Fard:2018lsf,Bottaro:2021snn,Bottaro:2022one,Bi:2024pkk}.

In this article we study the prospects of searching for LLPs in the context of the recently proposed $\mu$TRISTAN experiment\,\cite{Hamada:2022mua}. We focus in particular on its runs in the $\mu^+e^-$ mode with highly asymmetric beams $E_{\mu^+} \gg E_{e^-}$ that, despite achieving smaller center-of-mass (COM) energies with respect to the $\mu^+ \mu^+$ setup, will produce events highly boosted in the laboratory frame and concentrated within a small solid angle along the beam line. For LLPs produced in $\mu^+ e^-$ collisions, this feature can be exploited by placing a far detector along the beam line, which can therefore intercept a large fraction of the LLP flux.

We study this for the case of a long lived neutral scalar particle $\phi$ produced in pairs from the decay of the SM Higgs boson and decaying into a pair of SM particles, a common benchmark scenario used when comparing the reach of different LLPs searches. In particular, we show that a far detector with a volume comparable to the size of ANUBIS and placed  along the beam line at ${\cal O}(100)\,$m from the IP, would be capable, for specific $\phi$ decay modes, to extend the limits on the exotic branching ratio ${\rm BR}(h\to \phi\phi)$
beyond those expected from ATLAS and CMS searches at the end of the High Luminosity phase of the LHC (HL-LHC) in the large $c\tau_\phi$ region, where $c\tau_\phi$ is the $\phi$ proper decay length. On the other hand, we find that $\mu$TRISTAN will not be competitive with the bounds that might be set by the proposed LHC far detectors  CODEX-b, ANUBIS or MATHUSLA which, if realised, will extend well beyond its reach.\footnote{Similarly, the prospects for future lepton and hadron colliders are generally stronger than the ones attainable at $\mu$TRISTAN see, {\emph{e.g.}},\,\cite{Alipour-Fard:2018lsf,Bhattacherjee:2021rml}.}

The paper is organised as follows. In Sec.\,\ref{sec:muTristan} we briefly review the $\mu$TRISTAN project and discuss the Higgs boson production in  $\mu^+e^-$ mode. In Sec.\,\ref{sec:boosted_h} we study the production of boosted LLPs arising from the Higgs boson decay and present our sensitivity results in Sec.\,\ref{sec:results}. We then conclude in Sec.\,\ref{sec:conc}.

\section{$\mu$Tristan}\label{sec:muTristan}

$\mu$TRISTAN\,\cite{Hamada:2022mua} is a proposed experiment whose idea is to employ the method for cooling and focusing  positive muon beams used at J-PARC for the $(g-2)$ experiment\,\cite{Abe:2019thb} within a high-energy collider. Ideally, a low emittance $\mu^+$ beam could be accelerated up to energies of 1\,TeV and made it collide with 30\,GeV electrons from the SuperKEKB experiment in a 3\,km ring, resulting in a COM energy of $\sqrt s\simeq 346\,$GeV. The instantaneous luminosity estimate of $5\times 10^{33}\,{\rm cm}^{-2}\,{\rm s}^{-1}$ allows for an integrated dataset of ${\cal L}=1\,{\rm ab}^{-1}$ with ten years of data taking. A higher energy option with $E_{\mu^+}=3\,$TeV and $E_{e^-}=50\,$GeV in a 9\,km ring, corresponding to a COM energy of $\simeq 775\,$GeV and collecting a comparable data sample, might also be possible, assuming that 16\,T dipole magnets will be available by the time of construction\,\cite{Hamada:2022mua}. The possibility to have a polarised $e^-$ beam is also under consideration\,\cite{Roney:2021pwz}, while the $\mu^+$ beam 
will be highly polarised due to its production mode from $\pi^+$ decay. The machine can also run in $\mu^+ \mu^+$ mode with $\sqrt s = 2\,$TeV\,(6\,TeV) for the smaller (larger) ring size option, with an integrated luminosity of $\simeq 100\,$fb$^{-1}$ in both cases. 
All together both setups can be used to perform Higgs physics studies\,\cite{Hamada:2022mua}, SM precision measurements\,\cite{Hamada:2022uyn,Chen:2024tqh} as well as to probe NP up to the TeV scale\,\cite{Bossi:2020yne,Lu:2020dkx,Das:2022mmh,Dev:2023nha,Fridell:2023gjx,Fukuda:2023yui,Lichtenstein:2023iut,Das:2024gfg,Ding:2024zaj,Das:2024kyk,deLima:2024ohf,Sarkar:2025bgo}.

Regarding the Higgs boson, whose decay is the production mode for LLPs considered in this work, in $\mu^+ e^-$ mode it is mainly produced via charged and neutral vector boson fusion (VBF) processes
\be\label{eq:VBF}
\begin{split}
& \mu^+ e^- \to \bar\nu_\mu \nu_e h~~(WW~{\rm{fusion}}) \ , \\
& \mu^+ e^- \to \mu^+ e^- h~~(ZZ~{\rm{fusion}}) \ .
\end{split}
\ee
In this study we will consider beams with polarisation $({\cal P}_{\mu^+}, {\cal P}_{e^-}) =(+0.8, -0.7)$\,\cite{Hamada:2022mua} for which the total VBF cross section, computed at leading order (LO) with {\tt MadGraph5}\,\cite{Alwall:2014hca}, is $\sigma^{\rm VBF}_h \simeq 95.5\,{\rm fb}\,(493.7\,{\rm fb})$ for the low- and high-energy setup respectively. 
The event yield in the $\mu^+ e^-$ mode is larger with respect to the one that can be obtained in the $\mu^+\mu^+$ mode, which suffers from a smaller integrated luminosity and a smaller VBF cross section, only mediated by neutral $Z$ boson fusion.

As mentioned, the energy asymmetry of the $\mu^+e^-$ mode, hereafter $\mu$TRISTAN$_{\mu e}$, makes the final state particles of a collision event to be boosted towards the direction of the $\mu^+$ beam. For example, Higgs bosons produced in VBF have a distributions that peaks at a polar angle $\theta_h\simeq 5^\circ$. Also the Higgs boson decay products will be mostly concentrated close to the beam line, and a good detector coverage at small polar angles will thus be essential in order to perform Higgs precision measurements, possibly posing a challenge for detector design\,\cite{Hamada:2022mua}.

\section{Boosted Higgs decay}\label{sec:boosted_h}

The boosted nature of $\mu$TRISTAN$_{\mu e}$ can become an advantage when performing searches for LLPs promptly produced in $\mu^+e^-$ collisions and decaying into visible final states far
away from the IP.  The advantage arises because a far detector can only cover  a small portion of the total solid angle, of the order of $\simeq r^2/D^2$, where $r$ and $D$ are the typical detector transverse size and distance from the IP. However, if the LLP flux is concentrated in a small solid angle around the beam line, a detector with a well calibrated size and placement can intercept a large part of it and compensate with a larger geometrical acceptance the generically lower event yield of $\mu$TRISTAN$_{\mu e}$ compared to other experiments.\footnote{See also\,\cite{Fu:2022mtu,Barducci:2023gdc,deLima:2025ctj} for similar proposals in the context of different colliders.}

To illustrate this point, we consider the case of a beyond the SM  long lived neutral scalar particle $\phi$ produced in pairs from the decay 
of the SM Higgs boson, $h\to \phi\phi$, subsequently decaying into a pair of SM particles. Such decay can be produced, {\emph{e.g.}}, by a mixing in the scalar sector.  This is a standard benchmark scenario, commonly used to compare the sensitivity of different LLP search strategies. Such a signature can arise in several NP models, such as Higgs portal, Twin Higgs, Hidden Valley, supersymmetric extensions of the SM and multi-Higgs scenarios, see, {\emph{e.g.}},\,\cite{Curtin:2013fra} for a review.

Existing limits and prospects from LHC searches on this exotic decay mode depend on the LLP decay length. If $\phi$ decays displaced from the IP but within the detector volume, different search strategies can be adopted. Shorter decay lengths are tested via LLPs decaying in the ATLAS and CMS inner tracking systems\,\cite{ATLAS:2018rjc,CMS:2020iwv,ATLAS:2024qoo,CMS:2024xzb,CMS:2024qxz,CMS:2025qak}, while particles with larger decay lengths can be searched for through their decay in the ATLAS muon spectrometer\,\cite{ATLAS:2018tup,ATLAS:2019qrr,ATLAS:2019jcm}  
and the CMS end cap muon detector\,\cite{CMS:2021juv,CMS:2024bvl}.
These analyses allow to set limits down to  ${\rm BR}(h \to \phi\phi)\lesssim  10^{-3}-10^{-5}$, with the precise value depending on the nature of the final state products.\footnote{Experimental limits are normally presented in the $c\tau_\phi - {\rm BR}$ plane. What is, however, relevant for the effectiveness of the search is the laboratory frame decay length $\beta\gamma c\tau_\phi$, which depends on the $\phi$ boost, itself a function of both the $\phi$ mass and the production mechanism. Therefore, for a fixed production mode, limits for lighter $\phi$ shift towards smaller values of $c\tau_\phi$ compared to heavier $\phi$.} In particular, hadronic and di-muon decays are more strongly constrained with respect to decays into $e^+e^-$ and $\gamma\gamma$ final states. Conversely, if its decay length exceeds a value of ${\cal O}(10)\,$m, $\phi$ can remain invisible to ATLAS and CMS searches. Limits from the non observation of invisible Higgs decays set a limit of ${\rm BR}(h \to \phi\phi)\lesssim 15\%$\,\cite{ATLAS:2022yvh,CMS:2023sdw}, which will improve to ${\rm BR}(h \to \phi\phi)\lesssim 2\%$ by the end of the HL-LHC\,\cite{Cepeda:2019klc,ATL-PHYS-PUB-2025-018}. 

To test particles with very large decay lengths at the LHC, various proposals have been put forward. Typically, they involve large volume detectors placed far away from the IPs. In this way they also benefit from extremely low background rates and can then be sensitive to a handful of NP events in order to exclude or discover a decaying LLP. Far detectors can be on-axis, as the currently operating FASER\,\cite{FASER:2018eoc,Feng:2017uoz} or off-axis, as it is the case for the proposed experiments CODEX-b\,\cite{Gligorov:2017nwh,CODEX-b:2019jve,Dey:2019vyo}, ANUBIS\,\cite{Bauer:2019vqk,Brandt:2025fdj} and MATHUSLA\cite{Chou:2016lxi,Curtin:2018mvb,MATHUSLA:2018bqv,MATHUSLA:2025zyt}. If realised, these experiments will be able to test down to ${\rm BR}(h \to \phi\phi)\lesssim 10^{-5}-10^{-6}$ for large $c\tau_\phi$ value.

As mentioned, $\mu$TRISTAN$_{\mu e}$ is expected to produce a flux of $\phi$ particles from the decay of the SM Higgs boson with a small polar angle with respect to the beam line. 
To illustrate this, we generated $N_{\rm MC}=10^4$ parton level LO Monte Carlo (MC) events for the processes of Eq.\,\eqref{eq:VBF} via {\tt MadGraph5}. We then performed $N_{\rm toy}=10^6$ toy experiments, randomly sampling the Higgs boson four-momentum from the MC events. For each toy experiment, we built the  four-momentum of one of the two $\phi$s in the Higgs rest frame, $P^\mu_{\phi_0}$, where
\be
P^\mu_{\phi_0} = \frac{m_h}{2} \left( 1, \beta \sin\theta_{\phi_0}, 0, \beta \cos\theta_{\phi_0} \right),~\, \beta = \sqrt{1-\frac{4 m_\phi^2}{m_h^2}} \ .
\ee
Here $\theta_{\phi_0}$ is the $\phi$ polar angle in the Higgs boson rest frame, randomly sampled from a uniform distribution in $\cos\theta_{\phi_0}\in[-1,\,1]$,
given the isotropy of the  $h\to \phi\phi$ decay. We then boosted the $\phi$ four-momentum to the laboratory frame, $P^\mu_{\phi}$, and finally extracted the $\cos\theta_\phi$ distribution.

\begin{figure}[t!]
\begin{center}
\includegraphics[width=0.48\textwidth]{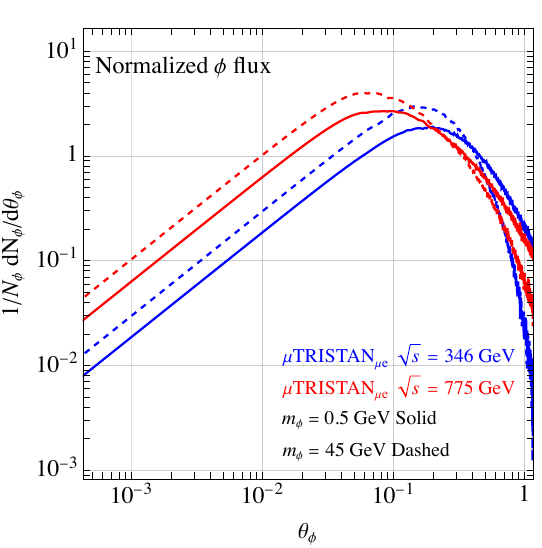}
\caption{\small Normalised flux of $\phi$ particles as a function of the polar angle $\theta_\phi$ for $m_\phi=0.5\,$GeV (solid lines) and $m_\phi=45\,$GeV (dashed lines) at $\mu$TRISTAN$_{\mu e}$ with $\sqrt s = 346\,$GeV (blue) and  $\sqrt s = 775\,$GeV (red).}
\label{fig:flux}
\end{center}
\end{figure}

The resulting normalized flux of $\phi$ particles for 
$\sqrt s=346\,$GeV and $\sqrt s=775\,$GeV and for different choices of $\phi$ masses, is illustrated in Fig.\,\ref{fig:flux}. 
We see that, at fixed COM energy, a heavier $\phi$ tends to be produced at smaller polar angles. This is to be expected, since the heavier the $\phi$s, the softer they are in the Higgs rest frame and therefore the boost to the laboratory frame tends to produce more collinear $\phi$ particles due to the asymmetric beams nature. On the other hand, at fixed mass, $\theta_\phi$ becomes smaller for higher $\sqrt s$, owing to the larger Lorentz boost of the Higgs boson. The peak of the distributions can be understood as follows. In the limit of massless $\phi$ and at fixed Higgs boson energy $E_h\gg |\vec p_h|$, assumed to be produced collinear to the beam line as expected in VBF processes, the polar angle in the laboratory frame is given by $\cos\theta_\phi = 1/\sqrt{1+\frac{1}{\gamma_h^2}\frac{1-\cos\theta_{\phi0}}{1+\cos\theta_{\phi0}}}$, where $\gamma_h = E_h/m_h$.  The ${\rm d}N_\phi / {\rm d}\theta_\phi$ distributions then shows a Jacobian peak located at $\theta_h \simeq 1/\sqrt{3\gamma_h^2-1}$, where the average $\gamma$ factors evaluated from the MC simulation are $\gamma_h \simeq 3.6\,(6.6)$ for the low- and high-energy setup respectively.

\begin{figure}[t!]
\begin{center}
\includegraphics[width=0.48\textwidth]{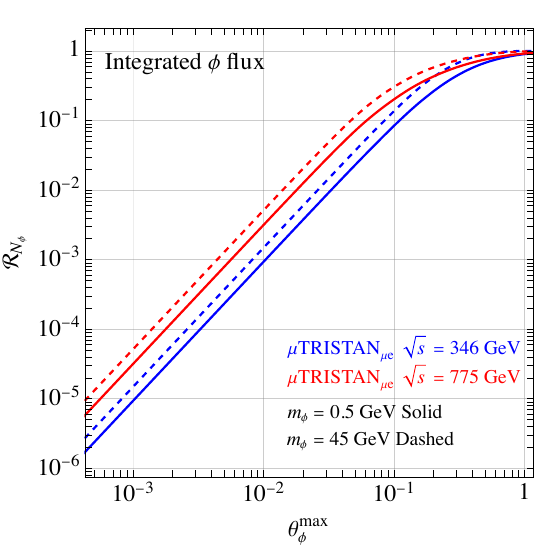}
\caption{\small 
Fraction of the total $\phi$ flux that can be intercepted assuming a cylindrical detector subtending an angle $\theta_{\rm max}$ for $m_\phi=0.5\,$GeV (solid lines) and $m_\phi=45\,$GeV (dashed lines) at $\mu$TRISTAN$_{\mu e}$ with $\sqrt s = 346\,$GeV (blue) and  $\sqrt s = 775\,$GeV (red).}
\label{fig:integrated_flux}
\end{center}
\end{figure}

The fraction ${\cal R}_{N_\phi}$ of the total $\phi$ flux that can be intercepted assuming, for simplicity, a cylindrical detector with radius $r$ placed along the beam line at a distance $D$ from the IP, thus subtending an angle $ \theta_\phi^{\rm max} = \tan^{-1}\left(\frac{r}{D}\right)$, is shown in Fig.\,\ref{fig:integrated_flux}. As an example, a detector of radius $r=10\,$m placed at a distance $D=100\,$m will intercept a flux of $\phi$ particles ${\cal R}_{N_\phi}\gtrsim 10\%$,  a factor $\simeq 40$ larger than the fraction of solid angle covered by the detector, $(1-\cos\theta_\phi^{\rm max})/2$.

\section{Expected sensitivity}\label{sec:results}

We estimate the expected sensitivity of $\mu$TRISTAN$_{\mu e}$ to long lived scalars arising from exotic Higgs boson decays through a MC simulation. Following the same procedure as in Sec.\,\ref{sec:boosted_h}, we generated $10^6$ events for $h\to \phi\phi$ decay. For each event, using the $\theta_{\phi_{1,2}}$ angles for both particles in the laboratory frame, we checked if the $\phi_{1,2}$ trajectory intersect the detector and, then, computed the probability for each particle to decay within the detector decay volume,  
assumed to be a cylinder of radius $r$ and length $L$ placed at a distance $D$ along the beam line. This probability is given by the exponential decay law
\be
{\cal P}(x_I,x_F) = e^{- \frac{x_I}{\beta\gamma c\tau_\phi}}-
e^{- \frac{x_F}{\beta\gamma c\tau_\phi}} \ ,
\ee
where, for our geometry,
\be
x_I = \frac{D}{\cos\theta_{\phi_i}} \, ,
\ee
and
\be
x_F = 
\begin{cases}
 \frac{D+L}{\cos\theta_{\phi_i}}~{\rm if}~~0 \le \tan\theta_{\phi_i} \le \frac{r}{D+L}\\
  \frac{r}{\sin\theta_{\phi_i}}~{\rm if}~~\frac{r}{D+L} \le \tan\theta_{\phi_i} \le\frac{r}{D}  
 \end{cases}
 \, ,
\ee
with $i=1,2$. Averaging over all the events we obtained the 
probability ${\cal P}_{\rm TOT}$ for at least one $\phi$ particle decaying in the detector volume. The total event yield is then computed as
\be
N_{\rm TOT} = \sigma_h^{\rm VBF} \times {\cal L} \times  {\rm BR}(h\to \phi\phi) \times \varepsilon \times {\cal P}_{\rm TOT} \ ,
\ee
where $\varepsilon$ is the efficiency for reconstructing the $\phi$ decay products, which we fix to be unity for any decay mode, as usually assumed for these types of projection studies.
The specific detector geometry, size and distance from the IP point, have been chosen in an approximate way as follows. We imposed a detector volume comparable to the one of the proposed experiment ANUBIS. Namely, $V=1.5\times V_{\rm ANUBIS}$ where $V_{\rm ANUBIS}\simeq 15\times 10^{3}\,$m$^3$\cite{Bauer:2019vqk}. Then for the low- and high-energy $\mu$TRISTAN$_{\mu e}$ options, which have different ring circumference sizes, we chose $r$ and $D$ in order to optimise the reach in the $c\tau_\phi - {\rm BR}(h\to \phi \phi)$ plane, under the constraints that the cylindrical detector of radius $r$ should not intercept the collider ring and that its length $L$ should not exceed  $\simeq 100\,$m. Practically, for the low-energy (high-energy) option we set $D=100\,$m $(150\,$m), $r=10\,$m $(7.5\,$m) and $L=70\,$m $(125\,$m). As regarding possible backgrounds to our proposed search, we note that all SM particles produced at the IP will either be  bent away by the main detector magnets or else stopped by the environment material before they have the change to reach the far detector, with the exception of muons and neutrinos. These are then the only particles that can induce background events by scattering within and/or upstream the detector volume. The muon induced background can be almost completely suppressed by employing a veto scintillator system as in FASER\,\cite{FASER:2018eoc}, while the neutrino flux from hadron decay is expected to be drastically reduced with respect to the LHC case (which for the LLP search at FASER\,\cite{FASER:2023tle} already gives a negligible contribution),
 since $\mu$TRISTAN employs lepton beams. At $\mu$TRISTAN there will however be a large flux of neutrinos arising from the decay of the circulating $\mu^+$ beam. With bunches of $\simeq 1.4\times 10^{10}$ muons injected with a frequency of 50\,Hz\,\cite{Hamada:2022mua} one expects around $7\times 10^{11}$ decaying muons per second. Only those decaying in the straight section of the collider ring close to the IP will however travel towards the far detector. Assuming a straight region of length $\simeq 100\,$m, this results in a flux of $\simeq 10^{10}\,{\rm sec}^{-1}$ of almost collinear neutrinos. With an operational time of 
$t_{\rm tot} = {\cal L_{\rm tot}}/{\cal L_{\rm ist}} = 1000\,{\rm fb^{-1}}/(4.6\times 10^{33}\,{\rm cm}^2\,{\rm sec}^{-1})\simeq 2.2\times 10^8\,{\rm sec}$\,\cite{Hamada:2022mua}, one obtains $4.4 \times 10^{18}$ neutrinos intersecting the proposed detector during the experiment lifetime. The number of neutrinos interacting with the detector volume can be roughly computed by using the interaction probability within the FASER volume which, from\,\cite{FASER:2019dxq,FASER:2023tle}, can be estimated to be around $4\times 10^{-16}$, rescaled by the different length of the proposed detector with respect to FASER. This gives approximately $10^5$ neutrino induced events within the detector volume, which is a non negligible rate.
Possible ways to suppress the neutrino induced background are {\emph{(i) Use of timing information:}} beam decay neutrinos come from continuous $\mu^+$ decay  while LLPs from hard scattering arrive in narrow
bunches. With precision timing it might be possible to reject events not coincident with
bunch collisions. {\emph{(ii) Track origin:}} LLPs decay into a pair of SM particles whose tracks originate in a common
vertex. A good reconstruction of both tracks can allow to discriminate signal events with
respect to neutrino induced background events.  {\emph{(iii) Improved detector design:}} minimising the amount of material in the main detector
volume can reduce the neutrino interaction rate. {\emph{(iv) Hollow cylindrical detector:} neutrinos have a tiny angular separation with respect to parent muon, see {\emph{e.g.}}\,\cite{Marzocca:2025inb}, while the $\phi$ flux peaks at a larger angle, see Fig.\,\ref{fig:flux}.  An hollow cylindrical
detector equipped with veto layers can allow to retain only tracks originating from LLPs decaying within
the main detector volume\,\footnote{To mitigate the neutrino induced background, one can also employ a slightly off-axis detector. This configuration also allows the detector to be placed closer to the IP, further benefiting from a reduced attenuation due to exponential decay.}.

Assuming that all the backgrounds can be reduced to a negligible level, we then computed the 95\% confidence level (CL) exclusion limits by requiring a maximum number of signal events $N_{\rm TOT}=4$. 

The results are illustrated in Fig.\,\ref{fig:res_1} for various choices of $m_\phi$. In the figures we also show, as shaded areas, the current limits arising  from LHC displaced vertices searches for specific $\phi$ decay modes and, with solid lines, their projected reach  at the end of the HL-LHC with an integrated luminosity of $3\,$ab$^{-1}$. We see that, despite the lower COM energy and Higgs production rate, a dedicated far detector at $\mu$TRISTAN$_{\mu e}$ can cover regions of the $c\tau_\phi - {\rm BR}(h\to \phi \phi)$ parameter space that will not be tested by displaced vertices searches. Although $\mu$TRISTAN$_{\mu e}$ cannot probe values of ${\rm BR}(h\to \phi \phi)$ below the minimum reachable at the end of the HL-LHC, it can improve on it in the  high $c\tau_\phi$ region, owing to the boosted nature of the Higgs boson and the large distance of the proposed detector with respect to the IP. In particular the low-energy run with $\sqrt s=346\,$GeV can extend over
 the HL-LHC projected limits for the case of  $\phi\to e^+ e^-,\,\tau^+\tau^-,\,\pi^{0}\pi^{0}\,\pi^\pm \pi^\mp$ and $\gamma\gamma$ decays for $m_\phi=0.5\,$GeV. The high-energy run with $\sqrt s=775\,$GeV can instead further extend the HL-LHC reach also in the case of hadronic decays for higher $\phi$ masses, while no improvement can be obtained in the case $\phi \to \mu^+\mu^-$ channel for $m_\phi > 15\,$GeV\,\footnote{Given the polar angular distribution shown in Fig.,\ref{fig:flux} which peaks at $\theta_\phi \simeq 0.1$, corresponding to $\eta_\phi \simeq 3$, one can also obtain a good sensitivity from the $\mu$TRISTAN main detectors located around the IP. 
We expect the proposed far detector to be complementary to the main ones, providing a better reach for large values of $c\tau$. A dedicated analysis of LLPs decaying within the main detectors is left for future work.}.
We also compare our results with the reach that could be obtained by the proposed LHC far detectors CODEX-b, ANUBIS and MATHUSLA\,\footnote{Also FASER\,\cite{FASER:2018eoc} and SND@LHC\,\cite{Boyarsky:2021moj} can be sensitive to light scalars with a trilinear coupling with the Higgs boson. However given that the main considered production mode in these experiments is via meson decay, their reach extends only up to $m_\phi \simeq {\cal O}({\rm GeV})$. For $m_\phi=0.5\,$GeV the reach of FASER2 doesn't extend beyond the projected reach of\,\cite{CMS:2024bvl} and of CODEX-b, while SND@LHC has weaker sensitivity.
}, represented as gray dashed lines.
Projections for these detectors generally assume the same sensitivity for any $\phi$ decay mode and we see that $\mu$TRISTAN$_{\mu e}$ will not be able to probe beyond the limits that will be set by these experiments, should they be realised at the HL-LHC. This is to be expected, owing to the very large LHC Higgs production cross section of $\simeq 50\,$pb and the proposed far detector geometries which, being 
off-axis, can be located closer to the IP, especially for the case of CODEX-b and ANUBIS.

\begin{figure*}[t!]
\begin{center}
\includegraphics[width=0.48\textwidth]{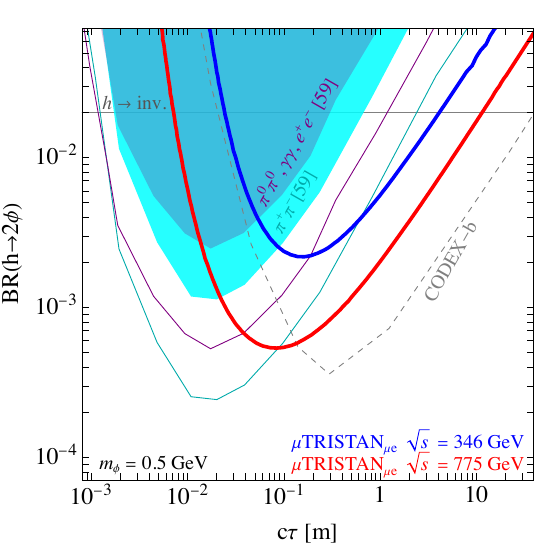}
\includegraphics[width=0.48\textwidth]{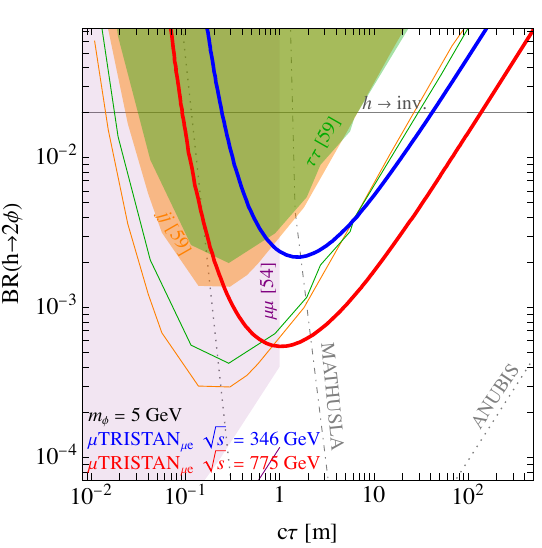}
\includegraphics[width=0.48\textwidth]{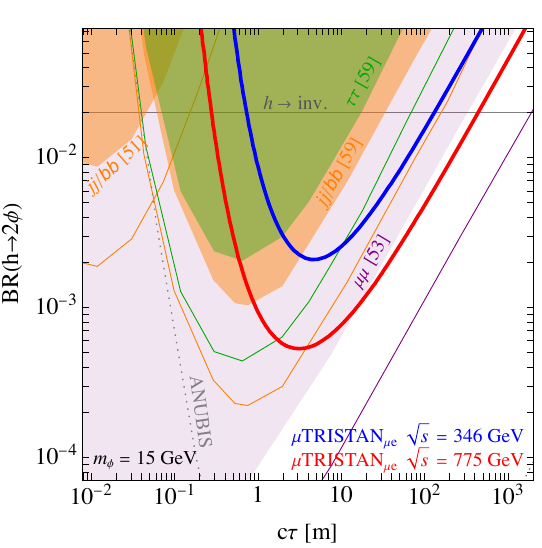}
\includegraphics[width=0.48\textwidth]{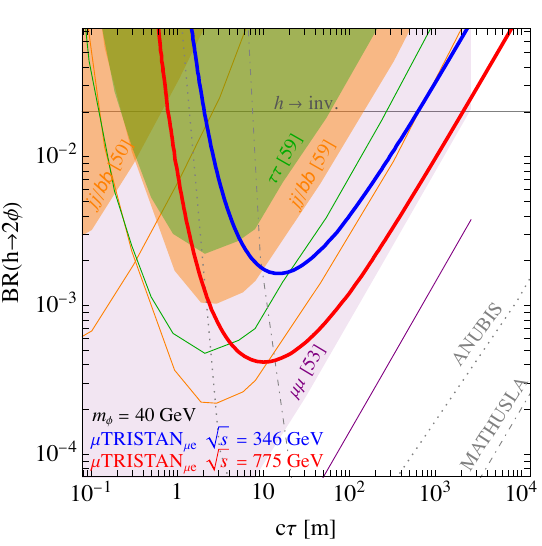}
\caption{\small 
Exclusion limits at 95\%~CL for the low-energy (blue) and high-energy (red) runs of $\mu$TRISTAN$_{\mu e}$ with a far detector along the beam line. We fix $D=100\,$m $(150\,$m), $r=10\,$m $(7.5\,$m) and $L=70\,$m $(125\,$m) for the low-energy (high-energy) runs respectively. The shaded areas represent current limits from LHC displaced vertices searches and the solid lines indicate their projected reach at the end of the HL-LHC with $3\,$ab$^{-1}$ of integrated luminosity. Also shown as an horizontal gray line is the limit on the invisible Higgs branching ratio expected with the same dataset. The dashed, dotted and dot-dashed lines represent the limits that could be obtained by the CODEX-b, ANUBIS and MATHUSLA experiments, respectively. }
\label{fig:res_1}
\end{center}
\end{figure*}

\section{Conclusions}\label{sec:conc}

The search for light and weakly coupled dark sectors is one of the priorities of the current LHC experimental program. Long lived particles with decay lengths of ${\cal O}(10^{-3}-10)\,$m can be searched for via displaced decays at ATLAS and CMS. On the other hand, particles with larger decay lengths, of ${\cal O}(10-100)\,$m can be within reach of proposed future far detectors, such as CODEX-b, ANUBIS and MATHUSLA. 
These elusive states are also a priority when forecasting the potential of future colliders for the post LHC era, and various studies have addressed this problem for  $e^+e^-$ colliders\,as FCC-ee, ILC and CEPC, hadron colliders as FCC-hh and  $\mu^+\mu^-$colliders.

In this work, we studied the prospects of the proposed $\mu$TRISTAN experiment running in the energy asymmetric $\mu^+ e^-$ mode for probing long lived particles arising from the exotic decay of the Standard Model Higgs boson. We considered the proposed low- and high-energy runs with $E_{\mu^+}=1\,$TeV and $E_{\mu^+}=3\,$TeV, corresponding to $\sqrt s\simeq346\,$GeV and $\sqrt s\simeq775\,$GeV, respectively.
Owing to the boosted nature of the events produced in the asymmetric collisions, a far detector placed along the beam line can collect a large fraction of the LLP flux. This allows setting limits on ${\rm BR}(h\to \phi\phi)$ that, for specific $\phi$ decay modes, can surpass those expected at the end of the HL-LHC in the large proper decay length part of the parameter space.
However, we  find that the proposed strategy is not competitive with the limits that might be set by currently proposed LHC far detectors such as CODEX-b, ANUBIS and MATHUSLA which, if realised, will enforce stronger limit that those attainable at $\mu$TRISTAN$_{\mu e}$.

\clearpage\newpage

\bibliography{biblio}

\end{document}